\documentclass[11pt]{article}
\usepackage{amsmath,amsthm,amssymb}
\date{}

\title{On the Covering Radius of the Second Order Reed-Muller Code of Length 128}
\author{Qichun Wang \footnote{Temasek Laboratories, National University of Singapore, 117411, Singapore,
 Email: tslwq@nus.edu.sg}}
\newtheorem{theorem}{Theorem}
\newtheorem{lemma}{Lemma}

\newtheorem{definition}{Definition}
\newtheorem{proposition}{Proposition}
\newtheorem{observation}{Observation}

\begin{document}
\maketitle
\begin{abstract}
In 1981, Schatz proved that the covering radius of the binary Reed-Muller code $RM(2,6)$ is 18. For $RM(2,7)$, we only know that its covering radius is between 40 and 44. In this paper, we prove that the covering radius of the binary Reed-Muller code $RM(2,7)$ is at most 42. Moreover, we give a sufficient and necessary condition for Boolean functions of 7-variable to achieve the second-order nonlinearity 42.
\end{abstract}


\section{Introduction}
In \cite{Schatz}, Schatz proved that the covering radius of the binary Reed-Muller code $RM(2,6)$ is 18. For $m\geq 7$, the covering radius of $RM(2,m)$ is still unknown. However, some bounds on it have been given \cite{Carlet1,Carlet2,Cohen}. From these bounds, we know that the covering radius of the binary Reed-Muller code $RM(2,7)$ is between 40 and 44.

In \cite{Fourquet}, the authors proposed an improved list decoding algorithm. Using that algorithm, they found a 7-variable Boolean function with the second order nonlinearity 38. In \cite{Burgess}, the authors studied the covering radius of binary Reed-Muller codes in the set of resilient Boolean functions, and found a 7-variable balanced Boolean function with the second order nonlinearity 40.

 In this paper, we prove that the covering radius of the binary Reed-Muller code RM(2,7) is at most 42. Moreover, we give a sufficient and necessary condition for Boolean functions of 7-variable to achieve the second-order nonlinearity 42.

 The paper is organized as follows. In Section~2, the necessary background is established. In Section~3, we give some observations which will be used afterwards. We then prove the main theorem in Section~4. We end in Section~5 with conclusions.
\section{Preliminaries}
Let $\mathbb{F}_{2}^{n}$ be the $n$-dimensional vector space over the
finite field $\mathbb{F}_{2}$. We denote by $B_{n}$ the set
of all $n$-variable Boolean functions, from $\mathbb{F}_{2}^{n}$ into $\mathbb{F}_{2}$.

Any Boolean function $f\in B_{n}$ can be uniquely represented
as a multivariate polynomial in
$\mathbb{F}_{2}[x_{1},\cdots,x_{n}]$,
\[
f(x_1,\ldots ,x_n)=\sum_{K\subseteq \{1,2,\ldots ,n\}}a_K\prod_{k\in K}x_k,
\]
 which is called its algebraic normal form
(ANF). The algebraic degree of $f$, denoted by $\deg(f)$, is the
number of variables in the highest order term with nonzero
coefficient.

 A Boolean function is {\em affine}
if there exists no term of degree strictly greater than 1 in the ANF. The set of all affine functions is denoted by~$A_{n}$.

 The cardinality of the set $\{x\in \mathbb{F}_{2}^{n}|f(x)=1\}$ is called the {\em Hamming weight} of $f$. The {\em Hamming distance} between two functions
$f$ and $g$ is the Hamming weight of $f+g$, and will be denoted by $d(f,g)$.

 Let $f\in B_{n}$. The {\em nonlinearity} of $f$ is its
distance from the set of all $n$-variable affine functions, that is,
\[
nl(f)=\min_{g\in A_{n}}d(f,g).
\]
The nonlinearity of an $n$-variable Boolean function is bounded above by $2^{n-1}-2^{n/2-1}$, and a function is said to be {\em bent} if it achieves this bound. It is known that the algebraic degree of a bent function is bounded above by $\frac{n}{2}$~\cite{Carlet0,cs09,Rothaus}.

The {\em $r$-order nonlinearity}, denoted by $nl_r(f)$, is its
distance from the set of all $n$-variable functions of algebraic degrees at most $r$.

The $r$-th order Reed-Muller code of length $2^n$ is denoted by $RM(r,n)$. Its codewords can be presented by the set of $n$-variable Boolean functions of degree $\leq r$. The {\em covering radius} of $RM(r,n)$ is defined as
\[
\max_{f\in B_n}d(f,RM(r,n))=\max_{f\in B_n}nl_r(f).
\]

We use $||$ to denote the concatenation. That is,
\[
f_1||f_2=(x_{n+1}+1)f_1+x_{n+1}f_2,
\]
 where $f_1,f_2\in B_n$.
\section{Some observations}
For $n=6$, classification of Boolean functions under the affine group has been fully studied (see e.g. \cite{Langevin,Maiorana}). It is known that there are exactly 205 affine equivalence classes modulo $RM(2,6)$. Investigating these affine equivalence classes, we have the following observations.
\begin{observation}
Let $f\in B_6$ and $f=x_1x_2x_3+x_1x_4x_5+x_2x_4x_6+x_3x_5x_6+x_4x_5x_6+g$, where $g$ is any 6-variable Boolean function with $\deg(g)\leq 2$. Then $nl(f)\leq 22$.
\end{observation}
\begin{observation}
Let $f\in B_6$. Then $nl_2(f)=17$ if and only if there is a $g\in B_6$ with $\deg(g)\leq 2$ such that $f$ is affine equivalent to $x_1x_2x_3x_4x_5x_6+x_1x_2x_3+x_1x_4x_5+x_2x_4x_6+x_3x_5x_6+x_4x_5x_6+g$.
\end{observation}
In the following, $fun_1=x_1x_2x_3+x_1x_4x_5+x_2x_4x_6+x_3x_5x_6+x_4x_5x_6$ and $fun_2=x_1x_2x_3x_4x_5x_6+fun_1$.
\begin{observation}
Let $f\in B_6$. Then $nl_2(f)=16$ if and only if there is a $g\in B_6$ with $\deg(g)\leq 2$ such that $f+g$ is affine equivalent to one of the following functions:

(1) $fun_3=x_1x_2x_6+x_1x_3x_5+x_2x_3x_4$;

(2) $fun_4=x_1x_2x_3x_4+x_1x_2x_6+x_1x_4x_5+x_2x_3x_5$;

(3) $fun_5=x_1x_2x_3x_4+x_1x_3x_5+x_1x_4x_6+x_2x_3x_5+x_2x_3x_6+x_2x_4x_5$;

(4) $fun_6=x_1x_2x_3x_6+x_1x_2x_4x_5+x_1x_3x_5+x_1x_4x_5+x_1x_4x_6+x_2x_3x_4$;

(5) $fun_7=x_1x_2x_3x_4x_5+x_1x_3x_5+x_1x_4x_6+x_2x_3x_5+x_2x_3x_6+x_2x_4x_5$.
\end{observation}
\begin{observation}
Let $f\in B_6$. Then $nl_2(f)=15$ if and only if there is a $g\in B_6$ with $\deg(g)\leq 2$ such that $f+g$ is affine equivalent to one of the functions $x_1x_2x_3x_4x_5x_6+fun_i$, where $3\leq i\leq 7$.
\end{observation}

\begin{definition}
Given $f\in B_n$, we denote by $Fh_f$ the map from $\mathbb{Z}$ to the power set of $B_n$ as follows:
\[
Fh_f(r)=\{g=\sum_{1\leq i<j\leq n}a_{ij}x_ix_j \ | \ a_{ij}\in \mathbb{F}_2 \ and \ nl(f+g)=r\}.
\]
We denote by $NFh_f$ the function from $\mathbb{Z}$ to $\mathbb{Z}$ whose value at $r$ is the cardinality of the set $Fh_f(r)$.
\end{definition}
Clearly, we have
\[
\sum_{i=0}^{\infty}NFh_f(i)=2^{n(n-1)/2}.
\]
Moreover, if $f_1$ is affine equivalent to $f_2$, then $NFh_{f_1}=NFh_{f_2}$.
\begin{observation}
We have

(1) $NFh_{fun_3}(16)=448$, $NFh_{fun_3}(26)=0$ and $NFh_{fun_3}(28)=64$;

(2) $NFh_{fun_4}(16)=384$, $NFh_{fun_4}(18)=1024$, $NFh_{fun_4}(20)=9216$, $NFh_{fun_4}(22)=14336$, $NFh_{fun_4}(24)=6784$, $NFh_{fun_4}(26)=10244$ and $NFh_{fun_4}(28)=0$;

(3) $NFh_{fun_5}(i)=0$, for $i\geq 26$;

(4) $NFh_{fun_6}(16)=224$, $NFh_{fun_6}(18)=1792$, $NFh_{fun_6}(20)=8640$, $NFh_{fun_6}(22)=14080$, $NFh_{fun_6}(24)=7520$, $NFh_{fun_6}(26)=512$ and $NFh_{fun_6}(28)=0$;

(5) $NFh_{fun_7}(i)=0$, for $i\geq 26$.
\end{observation}
\begin{observation}
We have $NFh_{x_1x_2x_3x_4x_5x_6+fun_i}(27)=0$, for $4\leq i\leq 7$. Moreover, $NFh_{fun_8}(15)=112$ and $NFh_{fun_8}(27)=64$, where $fun_8=x_1x_2x_3x_4x_5x_6+fun_3$.
\end{observation}
\begin{observation}
Let $f\in B_6$ and $nl_2(f)=14$. Then $NFh_{f}(r)=0$, for $r>26$. Moreover, if $NFh_{f}(26)>0$, then there is a $g\in B_6$ with $\deg(g)\leq 2$ such that $f+g$ is affine equivalent to one of the following functions:

(1) $fun_9=x_1x_2x_3x_4+x_1x_5x_6+x_2x_3x_6+x_2x_4x_5$; moreover, $NFh_{fun_9}(14)=16$ and $NFh_{fun_9}(16)=224$;

(2) $fun_{10}=x_1x_2x_3x_6+x_1x_2x_4x_5+x_1x_4x_5+x_1x_5x_6+x_2x_3x_5$; moreover, $NFh_{fun_{10}}(14)=32$ and $NFh_{fun_{10}}(16)=224$;

(3) $fun_{11}=x_1x_2x_3x_6+x_1x_2x_4x_5+x_1x_5x_6+x_2x_4x_6+x_3x_4x_5$; moreover, $NFh_{fun_{11}}(14)=16$ and $NFh_{fun_{11}}(16)=224$;

(4) $fun_{12}=x_1x_2x_5x_6+x_1x_3x_4x_6+x_2x_3x_4x_5+x_1x_2x_4+x_1x_3x_4+x_1x_3x_5+x_2x_3x_6$; moreover, $NFh_{fun_{12}}(14)=8$ and $NFh_{fun_{12}}(16)=224$;

(5) $fun_{13}=x_1x_2x_5x_6+x_1x_3x_4x_6+x_2x_3x_4x_5+x_1x_3x_4+x_1x_4x_5+x_2x_3x_6$; moreover, $NFh_{fun_{13}}(14)=24$ and $NFh_{fun_{13}}(16)=224$;

(6) $fun_{14}=x_1x_2x_3x_4x_5+x_1x_2x_6+x_1x_3x_5+x_2x_3x_4$; moreover, $NFh_{fun_{14}}(14)=48$ and $NFh_{fun_{14}}(16)=128$;

(7) $fun_{15}=x_1x_2x_3x_4x_5+x_1x_2x_5+x_1x_4x_6+x_2x_3x_6$; moreover, $NFh_{fun_{15}}(14)=24$ and $NFh_{fun_{15}}(16)=176$;

(8) $fun_{16}=x_1x_2x_3x_4x_5+x_1x_2x_3x_6+x_1x_2x_6+x_1x_3x_5+x_2x_3x_4$; moreover, $NFh_{fun_{16}}(14)=64$ and $NFh_{fun_{16}}(16)=160$;

(9) $fun_{17}=x_1x_2x_3x_4x_5+x_1x_2x_5x_6+x_1x_3x_4x_6+x_1x_2x_4+x_1x_3x_5+x_3x_4x_6$; moreover, $NFh_{fun_{17}}(14)=20$ and $NFh_{fun_{17}}(16)=224$;

(10) $fun_{18}=x_1x_2x_3x_4x_5+x_1x_2x_5x_6+x_1x_3x_4x_6+x_1x_2x_4+x_1x_3x_4+x_1x_3x_5+x_2x_5x_6+x_3x_4x_6$; moreover, $NFh_{fun_{18}}(14)=26$ and $NFh_{fun_{18}}(16)=212$;
\end{observation}
\noindent {\bf Remark 1.} From the above observations, it is easy to see that the maximum possible second-order nonlinearity of a 6-variable bent function is 16, and there is no 6-variable bent function with the second-order nonlinearity 14. Moreover, the function $x_1x_3x_4+x_1x_2x_5+x_1x_6+x_2x_4+x_3x_4+x_3x_5$ is a bent function with the second-order nonlinearity 16.
\section{A theorem on the covering radius of the binary Reed-Muller code RM(2,7)}
\begin{lemma} [\cite{Burgess}]
Let $f\in B_6$. Then $nl_2(f)=18$ if and only if there is a $g\in B_6$ with $\deg(g)\leq 2$ such that $f$ is affine equivalent to $fun_1+g$.
\end{lemma}
\begin{proposition}
Let $f\in B_7$ and $f=f_1||f_2$. If $nl_2(f)>40$, then $nl_2(f_1)\leq16$ and $nl_2(f_2)\leq16$.
\end{proposition}
\proof
Let $nl_2(f)>40$. We divide the proof into the following two cases.

{\em Case} 1: $nl_2(f_1)=18$ or $nl_2(f_2)=18$.

We assume without loss of generality that $nl_2(f_1)=18$. Then by Lemma 1, $f_1$ is affine equivalent to $fun_1+g_0$, where $g_0\in B_6$ and $\deg(g_0)\leq 2$. Therefore, by Observation 1, $nl(f_1+g_1)\leq 22$ for any $g_1\in B_6$ with $\deg(g_1)\leq 2$. Since $nl_2(f_2)\leq 18$, there exists a $g_2\in B_6$ with $\deg(g_2)\leq 2$ such that $d(f_2,g_2)\leq 18$. Since $nl(f_1+g_2)\leq 22$, there exists an $l\in B_6$ with $\deg(l)\leq 1$ such that $d(f_1,g_2+l)\leq 22$. Let $g=(g_2+l)||g_2=g_2+(x_7+1)l$. Then $nl_2(f)\leq d(f,g)\leq 40$. Hence, $nl_2(f_1)\leq 17$ and $nl_2(f_2)\leq 17$.

{\em Case} 2: $nl_2(f_1)=17$ or $nl_2(f_2)=17$.

We assume without loss of generality that $nl_2(f_1)=17$. Then by Observation 2, $f_1$ is affine equivalent to $fun_2+g_0$, where $g_0\in B_6$ and $\deg(g_0)\leq 2$. By Observation 1 and $d(fun_2,fun_1)=1$, we have $nl(f_1+g_1)\leq 23$ for any $g_1\in B_6$ with $\deg(g_1)\leq 2$. Since $nl_2(f_2)\leq 17$, there exists a $g_2\in B_6$ with $\deg(g_2)\leq 2$ such that $d(f_2,g_2)\leq 17$. Since $nl(f_1+g_2)\leq 23$, there exists an $l\in B_6$ with $\deg(l)\leq 1$ such that $d(f_1,g_2+l)\leq 23$. Let $g=(g_2+l)||g_2=g_2+(x_7+1)l$. Then $nl_2(f)\leq d(f,g)\leq 40$, and the result follows.
\endproof

\begin{lemma}
Let $f\in B_n$ and $f=f_1||f_2$. If
\[
NFh_{f_{i}}(n_2)>\sum_{k\geq n_1}NFh_{f_{j}}(k),
\]
 where $\{i,j\}=\{1,2\}$, then $nl_2(f)<n_1+n_2$.
\end{lemma}
\proof
We assume without loss of generality that $i=2$ and $j=1$. Since
\[
NFh_{f_2}(n_2)>\sum_{k\geq n_1}NFh_{f_1}(k),
 \]
 there exists a homogeneous polynomial $g_0\in B_{n-1}$ of degree 2 or 0 such that $nl(f_2+g_0)=n_2$ and $nl(f_1+g_0)<n_1$. That is, there are $l_1, l_2\in B_{n-1}$ with $\deg(l_1)\leq 1$ and $\deg(l_2)\leq 1$ such that $d(f_1+g_0+l_1)<n_1$ and $d(f_2+g_0+l_2)=n_2$. Let $g=(g_0+l_1)||(g_0+l_2)$. Then $d(f,g)<n_1+n_2$.
\endproof

\begin{proposition}
Let $f\in B_7$ and $f=f_1||f_2$. Let $nl_2(f_1)=nl_2(f_2)=16$. Then $nl_2(f)\leq 42$. Moreover, if $nl_2(f)=42$, then there exist $h_1, h_2\in B_6$ with $\deg(h_1)\leq 2$ and $\deg(h_2)\leq 2$ such that $f_1+h_1$ is affine equivalent to $fun_{i_1}$ and $f_2+h_2$ is affine equivalent to $fun_{i_2}$, where $i_1,i_2\in \{4,6\}$.
\end{proposition}
\proof
By Observation 3, there exist $g_1, g_2\in B_6$ with $\deg(g_1)\leq 2$ and $\deg(g_2)\leq 2$ such that $f_1$ is affine equivalent to $fun_i+g_1$ and $f_2$ is affine equivalent to $fun_j+g_2$, where $3\leq i,j \leq 7$. By Observation 5,
\[
NFh_{fun_j+g_2}(16)>NFh_{fun_i+g_1}(28).
\]
 Therefore, by Lemma 2 and $nl_2(f)$ is even, we have $nl_2(f)\leq 42$.
\\
Suppose $nl_2(f)=42$. Then $f_i$ ($i=1$ or 2) cannot affine equivalent to $fun_j+g$ ($j=5$ or 7) for any $g\in B_6$ of degree at most 2 (otherwise, by Observation 5, $nl_2(f)\leq 24+16=40$). Since
\[
NFh_{fun_3}(26)+NFh_{fun_3}(28)<NFh_{fun_j}(16)
\]
 for $j\in \{3,4,6\}$, by Lemma 2, $f_i$ ($i=1$ or 2) cannot affine equivalent to $fun_3+g$ for any $g\in B_6$ of degree at most 2, and the result follows.
\endproof
\begin{proposition}
Let $f\in B_7$ and $f=f_1||f_2$. If $nl_2(f_1)\leq 16$ and $nl_2(f_2)\leq 15$, then $nl_2(f)<42$.
\end{proposition}
\proof
We divide the proof into the following four cases.

{\em Case} 1: $nl_2(f_1)=16$ and $nl_2(f_2)=15$.

Suppose $nl_2(f)\geq 42$. Then by Observations 3-6, there exist $g_1, g_2\in B_6$ with $\deg(g_1)\leq 2$ and $\deg(g_2)\leq 2$ such that $f_1$ is affine equivalent to $fun_3+g_1$ and $f_2$ is affine equivalent to $fun_8+g_2$ (since $16+25=15+26<42$). Since
\[
NFh_{fun_8}(15)>NFh_{fun_3}(28)
\]
 and $nl_2(f)$ is odd, then by Lemma 2, $nl_2(f)\leq 41$.

{\em Case} 2: $nl_2(f_1)=16$ and $nl_2(f_2)=14$.

Suppose $nl_2(f)\geq 42$. Then by Observations 3 and 5, there exists a $g_1\in B_6$ with $\deg(g_1)\leq 2$ such that $f_1$ is affine equivalent to $fun_3+g_1$ (since $14+26<42$). Moreover, we have $NFh_{f_2}(26)>0$ (since $16+24<42$). Therefore, by Observations 7, there is a $g_2\in B_6$ with $\deg(g_2)\leq 2$ such that $f_2+g_2$ is affine equivalent to one of $fun_{i}$, where $9\leq i\leq18$. By Observations 5 and 7, it is easy to check that
\[
NFh_{fun_i}(16)>NFh_{fun_3}(26)+NFh_{fun_3}(28),
\]
 for $9\leq i\leq 18$. Hence, by Lemma 2, $nl_2(f)<42$.

{\em Case} 3: $nl_2(f_1)=15$ and $nl_2(f_2)=15$.

By Observations 4 and 6, we have $nl_2(f)\leq 15+27=42$. Moreover, if $nl_2(f)=42$, then there exist $h_1, h_2\in B_6$ with $\deg(h_1)\leq 2$ and $\deg(h_2)\leq 2$ such that $f_1+h_1$ and $f_2+h_2$ are affine equivalent to $fun_{8}$. Since
\[
NFh_{fun_8}(27)<NFh_{fun_8}(15),
\]
 then by Lemma 2, $nl_2(f)<42$.

{\em Case} 4: $nl_2(f_1)<15$ and $nl_2(f_2)<15$.

If $nl_2(f_1)\leq 13$ or $nl_2(f_2)\leq 13$, then $nl_2(f)\leq 13+28=41$. If $nl_2(f_1)=nl_2(f_2)=14$, then by Observations 7, we have $nl_2(f)\leq 14+26=40$.
\endproof
\noindent {\bf Remark 2.} Since $f_1||f_2$ is affine equivalent to $f_2||f_1$, if $nl_2(f_1)\leq 15$ and $nl_2(f_2)\leq 16$, we also have $nl_2(f)<42$.
\begin{theorem}
Let $f\in B_7$. Then $nl_2(f)\leq 42$. Moreover, $nl_2(f)=42$ if and only if the following conditions hold:

(1) $f$ is affine equivalent to
\[
f_1||f_2=fun_{i_1}||(fun_{i_2}(Ax+b)+g)
\]
 modulo $RM(2,7)$, where $i_1,i_2\in \{4,6\}$, $A\in {GL}_n(\mathbb{F}_2)$, $b\in \mathbb{F}_{2}^{n}$ and $g\in B_6$ are of degree at most 2.

(2) Moreover, for $\{i,j\}=\{1,2\}$, we have $Fh_{f_i}(16)\subseteq Fh_{f_j}(26)$, $Fh_{f_i}(18)\subseteq Fh_{f_j}(24)\cup Fh_{f_j}(26)$ and $Fh_{f_i}(20)\subseteq Fh_{f_j}(22)\cup Fh_{f_j}(24)\cup Fh_{f_j}(26)$.
\end{theorem}
\proof
By Propositions 1-3, we have $nl_2(f)\leq 42$. Moreover, if $nl_2(f)=42$, then there exist $h_1, h_2\in B_6$ with $\deg(h_1)\leq 2$ and $\deg(h_2)\leq 2$ such that
\[
f=(fun_{i_1}(A_1x+b_1)+h_1)||(fun_{i_2}(A_2x+b_2)+h_2),
\]
 where $i_1,i_2\in \{4,6\}$, $A_i\in {GL}_n(\mathbb{F}_2)$ and $b_i\in \mathbb{F}_{2}^{n}$, for $i=1,2$. Clearly,
 $f$ is affine equivalent to $f_1||f_2=fun_{i_1}||(fun_{i_2}(Ax+b)+g)$ modulo $RM(2,7)$, where $A=A_2A_1^{-1}$, $b=A_2A_1^{-1}b_1+b_2$ and $g=(h_1+h_2)(A_1^{-1}x+A_1^{-1}b_1)$. For $\{i,j\}=\{1,2\}$, suppose there is a function $g_1\in Fh_{f_i}(16)-Fh_{f_j}(26)$. Then $nl(f_i+g_1)=16$ and $nl(f_j+g_1)\leq 24$. Hence, there exist affine functions $l_1$ and $l_2$ such that $d(f_i+g_1,l_1)=16$ and $d(f_j+g_1,l_2)\leq 24$. Therefore, $nl_2(f)\leq 40$, which is a contradiction. Hence, $Fh_{f_i}(16)\subseteq Fh_{f_j}(26)$. Similarly, we have $Fh_{f_i}(18)\subseteq Fh_{f_j}(24)\cup Fh_{f_j}(26)$ and $Fh_{f_i}(20)\subseteq Fh_{f_j}(22)\cup Fh_{f_j}(24)\cup Fh_{f_j}(26)$.
 \\
 Let $q\in B_7$ be of degree at most 2. Then it can be written as $q_1||q_2$, where $q_1, q_2\in B_6$ have the same terms of degree 2. If the two conditions hold, then it is easy to check that $d(f,q)\geq 42$, and the result follows.
\endproof
\section{Conclusion}
In this paper, we prove that the covering radius of the binary Reed-Muller code $RM(2,7)$ is at most 42. Moreover, we give a sufficient and necessary condition for Boolean functions of 7-variable to achieve the second-order nonlinearity 42.
\def\refname{References}

\end{document}